\documentclass[12pt,a4paper]{article}
\usepackage{amsfonts,latexsym}
\usepackage{amsmath,amssymb}
\usepackage{graphicx,color}

\renewcommand{\thefootnote}{\fnsymbol{footnote}}

\begin{document}

\vspace{12mm}

\begin{center}
{{{\Large {\bf Scale-invariant power spectra from  a Weyl-invariant scalar-tensor theory}}}}\\[10mm]

{Yun Soo Myung$^a$\footnote{e-mail address: ysmyung@inje.ac.kr} and Young-Jai Park$^a$\footnote{e-mail address:yjpark@sogang.ac.kr}}\\[8mm]

{$^a$Institute of Basic Sciences and Department  of Computer Simulation, Inje University Gimhae 621-749, Korea\\[0pt]}
{$^b$Department of Physics, Sogang University, Seoul 121-742, Korea\\[0pt]}

\end{center}
\vspace{2mm}

\begin{abstract}
We  obtain scale-invariant scalar and tensor power spectra from a
Weyl-invariant scalar-tensor theory in de Sitter spacetime. This
implies that the Weyl-invariance guarantees  to implement the
scale-invariance of power spectrum in de Sitter spacetime. We
establish  a deep connection between the Weyl-invariance of the
action and scale-invariance of power spectrum in de Sitter
spacetime.

\end{abstract}
\vspace{5mm}



\vspace{1.5cm}

\hspace{11.5cm}{Typeset Using \LaTeX}
\newpage
\renewcommand{\thefootnote}{\arabic{footnote}}
\setcounter{footnote}{0}


\section{Introduction}

The conformal gravity $C^{\mu\nu\rho\sigma}C_{\mu\nu\rho\sigma}$ of
being invariant under the Weyl transformation of $g_{\mu\nu}\to
\Omega^2(x)\tilde{g}_{\mu\nu}$ has its own interests in quantum
gravity and cosmology~\cite{Mannheim:2005bfa}. Its appearance
related to the trace anomaly was established
in~\cite{BD,Riegert:1984zz}. Stelle~\cite{Stelle:1976gc} has first
introduced the quadratic curvature gravity  of
$a(R_{\mu\nu}^2-R^2/3)+b R^2$ to improve  the perturbatively
renormalizable property of Einstein gravity in Minkowski spacetime.
For the case of $ab\not=0$, the renormalizability was achieved but
the unitarity was violated. This means that  even though the
$a$-term  improves the ultraviolet divergence, it induces
simultaneously ghost excitations which spoil the unitarity.   This
issue is not resolved completely until now in Minkowski spacetime.

A purely conformal gravity implication to cosmological perturbation
was first studied in~\cite{Mannheim:2011is},  indicating that there
exists a difference between conformal and Einstein gravities in
their perturbed equations in  de Sitter (dS) spacetime. On later,
one of authors has computed an observable of tensor power
spectrum~\cite{Myung:2014cra,Myung:2015vya}, which is
scale-invariant during dS inflation. In the Einstein-Weyl gravity, the
role of Weyl-squared term was extensively studied in dS
spacetime~\cite{Clunan:2009er,Deruelle:2010kf,Deruelle:2012xv,Myung:2014jha}

On the other hand, the Lee-Wick scalar theory~\cite{Lee:1969fy,Lee:1970iw,Cai:2008qw}
whose action is given by
\begin{equation}\label{LW}
S_{\rm LW}=-\frac{1}{2}\int
d^4x\sqrt{-g}\Big[(\partial\phi)^2+\frac{1}{M^2}(\square\phi)^2\Big],
\end{equation}
where the scalar has a mass dimension 1 and $S_{\rm LW}$ is not
invariant under the Weyl transformations of $g_{\mu\nu}\to
\Omega^2(x)\tilde{g}_{\mu\nu}$ and $\phi \to \tilde{\phi}/\Omega$.
It has provided a scale-invariant scalar spectra when one requires
 $M^2=2H^2$ in dS spacetime~\cite{Myung:2014mla}.
Also, a fourth-order scalar theory with nonminimal derivative
coupling could induce a scale-invariant scalar spectra by requiring
that the nonminimal derivative coupling constant be $\xi=2/3$ in dS
spacetime~\cite{Myung:2015xha}.

 Hence, it is quite interesting to
find a proper scalar theory which may give us a scale-invariant
scalar power spectrum without introducing any artificial
adjustments. This might be  a desired Weyl-invariant scalar theory.
As is well-known, the simplest example of a Weyl-invariant theory is
a massless vector theory  described by $F^{\mu\nu}F_{\mu\nu}/4g^2$
and the massless Dirac equation is also Weyl-covariant.   In order
to obtain scale-invariant scalar and tensor spectra, one has to
combine this would-be scalar theory with conformal gravity, leading
a Weyl-invariant scalar-tensor theory.

In this work,  we propose two candidates for the Weyl-invariant
scalar-tensor theory. We will show that the Weyl-invariance
guarantees  to implement the scale-invariant power spectra
in dS spacetime, which are independent of wave number $k$.
This work establishes  a deep connection between the Weyl-invariance of the action and
scale-invariance of power spectrum in dS spacetime clearly.

\section{Weyl-invariant scalar-tensor theory }
In this  work we wish to explore   a deep connection between the
Weyl-invariance of the action and scale-invariance of power spectrum
in dS spacetime. In two dimensions the second-order scalar operator
$\square_2$ is covariant under the Weyl transformation with
$\Omega=e^\sigma$ in the sense that $\square_2\to e^{-2\sigma}
\tilde{\square}_2$. However, this is not true in other dimensions.
For example, one finds that $\square\to
e^{-2\sigma}[\tilde{\square}+2(\tilde{\nabla}^\mu\sigma
)\tilde{\nabla}_\mu]$ in four dimensions. Accordingly, the
fourth-order scalar operator $\square^2$ is not Weyl-covariant. In
order to make it Weyl-covariant, we need to introduce additional
terms like as $2R^{\mu\nu}\nabla_\mu\nabla_\nu-\frac{2}{3}R\square
+\frac{1}{3} \nabla^\mu R
\nabla_\mu$~\cite{Mottola:2010gp,Antoniadis:2011ib}.

Let us first consider a  Weyl-invariant scalar-tensor theory whose
action is given by fourth-order scalar theory and conformal gravity
\begin{equation} \label{WIST}
S_{\rm ST1}=\frac{1}{2}\int d^4x
\sqrt{-g}\Big[-(\square\phi)^2+2\Big(R_{\mu\nu}-\frac{R}{3}g_{\mu\nu}\Big)\partial^\mu\phi\partial^\nu\phi
-\frac{\alpha}{2}C^{\mu\nu\rho\sigma}C_{\mu\nu\rho\sigma}\Big]
\end{equation}
with $\alpha$ is a dimensionless coupling constant. The appearance
of the second term is necessary to have  the Weyl-invariant scalar
theory. Here the conformal gravity of Weyl-squared term is given by
\begin{eqnarray}
C^{\mu\nu\rho\sigma}C_{\mu\nu\rho\sigma}(\equiv
C^2)&=&2\Big(R^{\mu\nu}R_{\mu\nu}-\frac{1}{3}R^2\Big)+
(R^{\mu\nu\rho\sigma}R_{\mu\nu\rho\sigma}-4R^{\mu\nu}R_{\mu\nu}+R^2)
\end{eqnarray}
with the Weyl tensor
\begin{equation}
C_{\mu\nu\rho\sigma}=R_{\mu\nu\rho\sigma}-\frac{1}{2}\Big(
g_{\mu\rho}R_{\nu\sigma}-g_{\mu\sigma}R_{\nu\rho}-g_{\nu\rho}R_{\mu\sigma}+g_{\nu\sigma}R_{\mu\rho}\Big)+\frac{1}{6}R(g_{\mu\rho}g_{\nu\sigma}-g_{\mu\sigma}g_{\nu\rho}).
\end{equation}
Alternatively, the action (\ref{WIST}) can be rewritten as
\begin{equation} \label{WIST2}
S_{\rm ST2}=\frac{1}{2}\int d^4x
\sqrt{-g}\Big[-(\square\phi)^2+2\Big(G_{\mu\nu}+\frac{R}{6}g_{\mu\nu}\Big)\partial^\mu\phi\partial^\nu\phi
-\frac{\alpha}{2}C^2\Big],
\end{equation}
where $G_{\mu\nu}\partial^\mu\phi\partial^\mu\phi$ denotes the
nonminimal derivative coupling term~\cite{Sushkov:2009hk} which may
render slow-roll inflation even for a steep
potential~\cite{Germani:2010gm,Germani:2011ua}. Here
$G_{\mu\nu}=R_{\mu\nu}-(R/2)g_{\mu\nu}$ is the Einstein tensor. This
expression shows clearly why (\ref{WIST}) differs from the
fourth-order scalar theory with nonminimal derivative coupling
model~\cite{Myung:2015xha}.

Noting that the Weyl-squared term is covariant
($C^2=e^{-4\sigma}\tilde{C}^2$) under the Weyl transformation of
$g_{\mu\nu} \to e^{2\sigma(x)}
\tilde{g}_{\mu\nu}$~\cite{Riegert:1984kt}, the first-two terms of
(\ref{WIST}) can be expressed as
\begin{equation}
\frac{1}{2}\int d^4x \sqrt{-g}\phi \Delta_4 \phi.
\end{equation}
Here the Weyl operator $\Delta_4$ takes the form
\begin{equation}
\Delta_4=\square^2+2R^{\mu\nu}\nabla_\mu\nabla_\nu-\frac{2}{3}R\square
+\frac{1}{3} \nabla^\mu R \nabla_\mu,
\end{equation}
which is obviously  Weyl-covariant
($\Delta_4=e^{-4\sigma}\tilde{\Delta}_4$) under the Weyl
transformation.  It is clear that  the action (\ref{WIST}) is
Weyl-invariant provided that the scalar field is dimensionless, when
one takes into account $\sqrt{-g}\to e^{4\sigma}\sqrt{-\tilde{g}}$.
Hence, the action (\ref{WIST}) is regarded as a promisingly
Weyl-invariant scalar-tensor theory, in compared to the Lee-Wick
scalar theory (\ref{LW}) and the fourth-order scalar theory with
nonminimal derivative coupling.

Now, we derive  the  Einstein  equation from (\ref{WIST})
\begin{equation} \label{ein-eq}
-\alpha B_{\mu\nu}=T_{\mu\nu},
\end{equation}
where the Bach tensor is defined  by
\begin{eqnarray} \label{equa2}
B_{\mu\nu}&=&2R^{\rho\sigma}(R_{\mu\rho\nu\sigma}-\frac{1}{4}
R_{\rho\sigma}g_{\mu\nu})-\frac{2}{3} R\Big(R_{\mu\nu}-\frac{1}{4}
Rg_{\mu\nu}\Big)\nonumber \\
&+&
\nabla^2R_{\mu\nu}-\frac{1}{6}\nabla^2Rg_{\mu\nu}-\frac{1}{3}\nabla_\mu\nabla_\nu
R.
\end{eqnarray}
Here  $T_{\mu\nu}$ is  the total energy-momentum tensor derived from
the first-three terms of (\ref{WIST}) which takes the
form~\cite{Mottola:2010gp}
\begin{eqnarray}
T_{\mu\nu}&=&-2\, \nabla_{(\mu}\phi \nabla_{\nu)} \square \phi +
2\,\nabla^\rho(\nabla_\rho \phi\nabla_\mu\nabla_\nu\phi) -
\frac{2}{3}\, \nabla_\mu\nabla_\nu(\nabla_\rho \phi
\nabla^\rho\phi)\nonumber\\
&& + \frac{2}{3}\,R_{\mu\nu}\, \nabla_\rho \phi\nabla^\rho \phi -
4\, R_{\rho(\mu}\nabla_{\nu)} \phi \nabla^\rho \phi
+ \frac{2}{3}\,R \,\nabla_\mu \phi\nabla_\nu \phi\label{cem-ten}\\
&& + \frac{1}{6}\, g_{\mu\nu}\, \left\{-3\, (\square\phi)^2 +
\square(\nabla_\rho\phi\nabla^\rho\phi) + 2\left( 3R^{\rho\sigma} -
R g^{\rho\sigma} \right) \nabla_\rho\phi\nabla_\sigma \phi\right\}.
\nonumber
\end{eqnarray}
On the other hand, its scalar equation is given by
\begin{equation}
\Delta_4 \phi=0.
\end{equation}
For a maximally symmetric spacetime with $\bar{R}$=const, one finds
\begin{equation}
\bar{R}_{\mu\nu\rho\sigma}=\frac{\bar{R}}{12}\Big(\bar{g}_{\mu\rho}\bar{g}_{\nu\sigma}-\bar{g}_{\mu\sigma}\bar{g}_{\nu\rho}\Big),~\bar{R}_{\mu\nu}=\frac{\bar{R}}{4}\bar{g}_{\mu\nu}.
\end{equation}
In this case, the Bach tensor is always zero ($\bar{B}_{\mu\nu}=0$).
Hence, choosing $\bar{\phi}$=const ($\bar{T}_{\mu\nu}=0$), we have
solutions of dS  ($\bar{R}>0$), Minkowski ($\bar{R}=0$), and anti de
Sitter ($\bar{R}<0$) spacetime. In this work, we concentrate on
 the dS solution for cosmological implication,  whose  curvature quantities
are given by
\begin{equation}
\bar{R}_{\mu\nu\rho\sigma}=H^2(\bar{g}_{\mu\rho}\bar{g}_{\nu\sigma}-\bar{g}_{\mu\sigma}\bar{g}_{\nu\rho}),~~\bar{R}_{\mu\nu}=3H^2\bar{g}_{\mu\nu},~~\bar{R}=12H^2
\end{equation}
with $H$=const.

 Now, let us  choose  dS  background explicitly  by
choosing a conformal time $\eta$
\begin{eqnarray} \label{frw}
ds^2_{\rm dS}=\bar{g}_{\mu\nu}dx^\mu
dx^\nu=a(\eta)^2\Big[-d\eta^2+d{\bf x}\cdot d{\bf x}\Big],
\end{eqnarray}
where the conformal  scale factor is
\begin{eqnarray}
a(\eta)=-\frac{1}{H\eta},
\end{eqnarray}
while the cosmic scale factor  is given by  $a(t)=e^{Ht}$ in a flat
Friedmann-Robertson-Walker (FRW) background $ds^2_{\rm FRW}=-dt^2+
a^2(t)d{\bf x}\cdot d{\bf x}$. We note that the dS solution is not
distinctive since any maximally symmetric spacetime can be a
solution  to Einstein and scalar equations. This redundancy of
solutions is a feature of Weyl-invariant scalar-tensor theory
(\ref{WIST}). The dS SO(1,4)-invariant distance between two
spacetime points $x^\mu$ and $x'^{\mu}$ is defined  by
\begin{eqnarray}
\label{id2} {\cal
Z}(x,x')=1-\frac{-(\eta-\eta')^2+|{\bf{x}}-{\bf{x}'}|^2}{4\eta\eta'}=1-\frac{(x-x')^2}{4\eta\eta'}
\end{eqnarray}
since ${\cal Z}(x,x')$ has the ten symmetries which leave the metric
of dS spacetime invariant. Here $(x-x')^2$ is the Lorentz-invariant
flat spacetime distance.

At this stage, it seems appropriate to comment on  the other
Weyl-invariant scalar-tensor theory of massive conformal
gravity~\cite{Faria:2013hxa,Myung:2014aia}
\begin{equation} \label{WIST3}
S_{\rm ST3}=-\frac{1}{2}\int d^4x
\sqrt{-g}\Big[(\partial\phi)^2+\frac{1}{6}\phi^2 R
+\frac{\alpha}{2}C^{\mu\nu\rho\sigma}C_{\mu\nu\rho\sigma}\Big],
\end{equation}
where its Weyl-invariance can be achieved  up to surface terms  by
requiring both $\phi \to \tilde{\phi}e^{-\sigma}$ and $g_{\mu\nu}
\to e^{2\sigma} \tilde{g}_{\mu\nu}$. Hence,  we wish to point out a
difference between ST1~\cite{Mottola:2010gp,Antoniadis:2011ib} and
ST3: the scaling dimension of $\phi$ in (\ref{WIST}) is zero, while
the scaling dimension of $\phi$ in (\ref{WIST3}) is $-1$ (or mass
dimension 1) as $\phi$ in kinetic term of $\phi \square \phi$ does
have. Also, $\phi$ in  (\ref{WIST}) is Weyl-invariant ($\phi\to
\tilde{\phi}$), whereas  $\phi$ in (\ref{WIST3}) transforms as $\phi
\to \tilde{\phi}e^{-\sigma}$.  Furthermore, since (\ref{WIST3})
provides a conformal scalar propagation in dS spacetime, it is not a
promising candidate for our purpose.

Adapting  the action (\ref{WIST3}) to find the background solution,
one finds Einstein and scalar equations
\begin{eqnarray}
\label{eqwi3-1}&&-\alpha B_{\mu\nu}=T^{\phi}_{\mu\nu}, \\
\label{eqwi3-2}&&\Big(\square-\frac{R}{6}\Big)\phi=0,
\end{eqnarray}
where
\begin{equation}
T^{\phi}_{\mu\nu}=\nabla_\mu\phi\nabla_\mu\phi
-\frac{1}{2}(\nabla\phi)^2g_{\mu\nu}+\frac{1}{6}G_{\mu\nu}\phi^2-\frac{1}{6}(\nabla_\mu\nabla_\nu-g_{\mu\nu}\square)\phi^2.
\end{equation}
Its trace is zero when using (\ref{eqwi3-2}). For
$\bar{\phi}=$const, the Minkoswksi  spacetime of
$\bar{R}=0(\bar{G}_{\mu\nu}=0)$ is only a solution. In  case of
$\bar{\phi}=0$, any maximally symmetric spacetime is a solution and
thus, dS solution is not distinctive.

Finally, if one wishes really to obtain dS  solution, one has to
insert a term of $R-2\Lambda(\Lambda=3H^2)$ into the action
(\ref{WIST}) which breaks the  Weyl-invariance manifestly. This
leads to a fourth-order scalar theory coupled to Einstein-Weyl
gravity, where one could not obtain a scale-invariant tensor
spectrum.

\section{Perturbed equations on de Sitter spacetime}
In order to derive  perturbed equation (linearized equation) around
dS spacetime, we introduce a perturbed scalar $\varphi$ as
\begin{equation}
\phi=\bar{\phi}+\varphi.
\end{equation}
For a metric perturbation,
 we   choose  the Newtonian gauge~\cite{Mukhanov:1990me}
of $B=E=0 $ and $\bar{E}_i=0$, leading  to $10-4=6$ degrees of
freedom (DOF). In this case, the cosmologically perturbed metric can
be simplified to be
\begin{eqnarray} \label{so3-met}
ds^2=a(\eta)^2\Big[-(1+2\Psi)d\eta^2+2\Psi_i d\eta
dx^{i}+\Big\{(1+2\Phi)\delta_{ij}+h_{ij}\Big\}dx^idx^j\Big]
\end{eqnarray}
with the transverse vector $\partial_i\Psi^i=0$ and
transverse-traceless tensor $\partial_ih^{ij}=h=0$. It is worth to
note that choosing the SO(3)-perturbed metric (\ref{so3-met})
contrasts  with the covariant approach to the  cosmological
conformal gravity~\cite{Mannheim:2011is}.

In order to get  the cosmological perturbed equations,  one  is
first to obtain the bilinear action and then, varying it to yield
the linearized equations. According to the previous
work~\cite{Deruelle:2010kf},   we expand the Weyl-invariant
scalar-tensor action (\ref{WIST}) up to quadratic order in the
perturbations of $\varphi,~\Psi,~\Phi,~\Psi_i,$ and $h_{ij}$ around
the dS  background.  Then,  the bilinear action is composed of four
terms as
\begin{equation}
\delta S_{\rm ST1}=\delta S_{\rm S}+\delta S_{\rm CG}^{({\rm S})}+
\delta S_{\rm CG}^{({\rm V})}+\delta S_{\rm CG}^{({\rm T})},
\end{equation}
where
\begin{eqnarray}
&&\hspace*{-2.3em}\delta S_{\rm S}=\frac{1}{2}\int d^4x~\varphi
\bar{\Delta}_4 \varphi,
\label{scalar0}\\
&&\hspace*{-2.3em}\delta S_{\rm CG}^{({\rm S})}=\frac{\alpha}{3}\int
d^4x\Big[\nabla^2 (\Psi-\Phi)\Big]^2,
\label{scalar}\\
&&\nonumber\\
 &&\hspace*{-2.3em}\delta S_{\rm CG}^{({\rm V})}=\frac{\alpha}{4}\int
d^4x\Big(\partial_i\Psi'_{j}\partial^{i}\Psi'^{j}
-\nabla^2\Psi_i\nabla^2\Psi^i\Big),\label{vpeq}\\
&&\nonumber\\
&&\hspace*{-2.3em} \delta S_{\rm CG}^{({\rm
T})}=\frac{\alpha}{8}\int d^4x\Big(h''_{ij}h''^{ij}
-2\partial_kh'_{ij}\partial^{k}h'^{ij}
+\nabla^2h_{ij}\nabla^2h^{ij}\Big)\label{hpeq}.
\end{eqnarray}
Here $'$(prime) denotes differentiation with respect to conformal time $\eta$.
Note that all of bilinear actions are independent of conformal scale
factor $a(\eta)$, showing that the Weyl-invariance persists in the
bilinear action.

 From (\ref{scalar0}), we obtain the fourth-order  perturbed scalar equation
\begin{equation} \label{varp-eq}
\bar{\Delta}_4 \varphi=-\Box(-\Box+2H^2)\varphi=0,
\end{equation}
which shows a second factorization of $\bar{\Delta}_4$ into two
second-order operators  in dS spacetime (and in fact any conformally
flat spacetime).  Here $\Box=-d^2/d\eta^2+\nabla^2$ with
$\nabla^2=\partial_i^2$ the Laplacian operator.
 Varying (\ref{vpeq}) and (\ref{hpeq}) with respect
to $\Psi^{i}$ and $h^{ij}$  leads to linearized  equations of motion
for vector and tensor perturbations
\begin{eqnarray}
&&\nabla^2 \Box\Psi_i=0,\label{veq}\\
 &&\Box^2h_{ij}=0.\label{heq}
\end{eqnarray}
 It is emphasized again that   (\ref{varp-eq})-(\ref{heq})  are independent
 of $a^2(\eta)$ of expanding dS background in the Weyl-invariant  scalar-tensor theory.

Finally, we would like to mention two scalars $\Phi$ and $\Psi$.
Two scalar equations are given by $\nabla^2\Psi=\nabla^2\Phi=0$,  which imply
that they are obviously
 non-propagating modes  in the dS background. Hereafter,
  we will not consider these irrelevant  metric-scalars.
  This means that the Weyl-invariant theory (\ref{WIST})  describes 7 DOF (1 scalar+ 2 of vector +4 of tensor
  modes),
  where the last becomes four because $h_{ij}$ satisfies a fourth-order
  equation.

\section{Primordial power spectra}
The power spectrum is usually  given by the two-point correlation
function which could be  computed when one chooses   the vacuum
state $|0\rangle$. It is defined by
\begin{equation} \label{ps1}
\langle0|{\cal F}(\eta,\bold{x}){\cal
F}(\eta,\bold{x}')|0\rangle=\int d^3\bold{k} \frac{{\cal P}_{\cal
F}(\eta,k)}{4\pi k^3}e^{i \bold{k}\cdot (\bold{x}-\bold{x}')},
\end{equation}
where ${\cal F}$ denotes scalar, vector or tensor, and
$k=|\bold{k}|$ is the wave number. For simplicity, we may use the
zero-point correlation function to define the power spectrum
as~\cite{Baumann:2009ds}
\begin{equation} \label{ps2}
\langle0|{\cal F}(\eta,0){\cal F}(\eta,0)|0\rangle=\int \frac{dk}{k}
{\cal P}_{\cal F}(\eta,k).
\end{equation}

In general, fluctuations are created on all length scales with wave
number $k$. Cosmologically relevant fluctuations start their lives
inside the Hubble radius which defines the subhorizon: $k~\gg
aH~(z=-k\eta\gg 1)$.  On the other hand, the comoving Hubble radius
$(aH)^{-1}$ shrinks during inflation while the comoving wave number
$k$ is constant. Therefore, eventually all fluctuations exit the
comoving  Hubble radius, which defines the superhorizon: $k~\ll
aH~(z=-k\eta\ll 1)$. One may compute the two-point function by
taking the Bunch-Davies vacuum $|0\rangle$.
 In the dS inflation, we choose the subhorizon limit
of  $z\to \infty$  to define the Bunch-Davies vacuum, while we
choose the superhorizon limit of $z\to 0$ to get a definite form of
the power spectrum which stays alive after decaying.

\subsection{Scalar power spectrum}
There are two ways to obtain  the scalar power spectrum: One is to
find the inverse Weyl operator $\bar{\Delta}_4^{-1}$ and
Fourier-transforming it leads to the scalar power spectrum. The
other is to compute the power spectrum (\ref{ps2}) directly by using
the quantization scheme of the  non-degenerate Pais-Uhlenbeck (PU)
oscillator. We briefly describe both computation schemes.

The inverse Weyl-operator is given by
~\cite{Mottola:2010gp,Antoniadis:2011ib}
\begin{equation} \label{4thp}
\bar{\Delta}_4^{-1}[{\cal
Z}(x,x')]=\frac{1}{2H^2}\Big[\frac{1}{-\Box}-\frac{1}{-\Box+2H^2}\Big]
=\frac{1}{2H^2}[G_{\rm mmc}[{\cal Z}(x,x')]-G_{\rm mcc}[{\cal
Z}(x,x')]],
\end{equation}
where the propagators of massless minimally coupled (mmc)
scalar~\cite{Bros:2010wa} and massless conformally coupled (mcc)
scalar~\cite{Higuchi:2009ew} in dS spacetime are given by
\begin{equation}\label{prop}
G_{\rm mmc}[{\cal
Z}(x,x')]=\frac{H^2}{(4\pi)^2}\Big[\frac{1}{1-{\cal Z}}-2\ln(1-{\cal
Z})+c_0\Big],~~G_{\rm mcc}[{\cal
Z}(x,x')]=\frac{H^2}{(4\pi)^2}\frac{1}{1-{\cal Z}},
\end{equation}
where the former is the dS invariant renormalized two-point function
(on the space of non-constant modes), while the latter is the
conformally  coupled scalar two-point function on dS spacetime. As
opposed to Ref.~\cite{Anderson:2013zia}, the inverse Weyl-operator
(\ref{4thp}) is dS-invariant because it is a function of $1-{\cal
Z}$. Substituting (\ref{prop}) into (\ref{4thp}), the propagator
takes the form
\begin{equation} \label{log-pro}
\bar{\Delta}_4^{-1}[{\cal
Z}(x,x')]=\frac{1}{16\pi^2}\Big(-\ln[1-{\cal
Z}(x,x')]+\frac{c_0}{2}\Big),
\end{equation}
 which is a purely logarithm up to an additive constant
 $c_0$ and is a dS-invariant two-point function.
The scalar power spectrum is defined by Fourier transforming
 the propagator at equal time $\eta=\eta'$ as
 \begin{eqnarray}  {\cal P}_\varphi&=&\frac{1}{(2\pi)^3}\int
d^3{\bf r} ~4\pi k^3 \bar{\Delta}_4^{-1}[{\cal Z}(\eta,{\bf
x};\eta,{\bf x}')]e^{-i{\bf k}\cdot {\bf
r}},~~{\bf r}={\bf x}-{\bf x}'\label{cesa1}\\
&=&\frac{1}{(2\pi)^3}\frac{k^3}{4\pi}\int d^3{\bf r}
\Big(-\ln\Big[\frac{r^2}{4\eta^2}\Big]+\frac{c_0}{2}\Big)e^{-i{\bf
k}\cdot {\bf
r}}\label{cesa2}\\
&=&-\frac{k^2}{8\pi^3}\int_0^{\infty} dr
\Big\{r\sin[kr]\ln[r^2]\Big\} \label{cesa4}\\
&=&\frac{1}{8\pi^2}, \label{cesa5}
\end{eqnarray}
where we have used the  Ces\`{a}ro-summation method in deriving from
(\ref{cesa4}) to (\ref{cesa5})~\cite{Youssef:2012cx,Myung:2015xha}.

On the other hand, Eq.(\ref{varp-eq}) implies two second-order
equations for mmc and mcc scalars
\begin{eqnarray}
\Box\varphi^{{\rm mmc}}&=&0,\label{eeq}\\
(\Box-2H^2)\varphi^{{\rm mcc}}&=&0\label{meq}.
\end{eqnarray}
Expanding $\varphi^{\rm mmc,mcc}$ in terms of Fourier modes
$\phi^{\rm mmc,mcc}_{\bf k}(\eta)$
\begin{eqnarray}\label{sfour}
\varphi^{\rm mmc,mcc}(\eta,{\bf
x})=\frac{1}{(2\pi)^{\frac{3}{2}}}\int d^3{\bf k}~\phi^{\rm
mmc,mcc}_{\bf k}(\eta)e^{i{\bf k}\cdot{\bf x}},
\end{eqnarray}
With $z=-k\eta$, Eqs.(\ref{eeq}) and (\ref{meq}) become
\begin{eqnarray}\label{s-eq2}
\Big(\frac{d^2}{d z^2}-\frac{2}{z}\frac{d}{d z}+1\Big)\phi^{\rm
mmc}_{\bf
k}&=&0,\label{pmsoll}\\
\Big(\frac{d^2}{dz^2}-\frac{2}{z}\frac{d}{d
z}+1+\frac{2}{z^2}\Big)\phi^{\rm mcc}_{\bf k}&=&0.\label{pmmsoll}
\end{eqnarray}
 Solutions to (\ref{pmsoll}) and (\ref{pmmsoll})
are given by
\begin{eqnarray}
\phi_{\bf k}^{\rm mmc}&=&c_{\rm mmc}(i+z)e^{iz},\label{pmsoll1}\\
\phi_{\bf k}^{\rm mcc}&=&c_{\rm mcc}ize^{iz},\label{pmmsoll1}
\end{eqnarray}
where $c_{\rm mmc}$ and $c_{\rm mcc}$ are constants to be
determined. Then, the field operator $\hat{\varphi}$ can be expanded in
Fourier modes as
\begin{eqnarray}\label{phiex}
\hat{\varphi}(\eta,{\bf x})=\frac{1}{(2\pi)^{\frac{3}{2}}}\int
d^3{\bf k}\Big[\Big(\hat{a}_1({\bf k})\phi_{\bf k}^{\rm
mmc}(\eta)+\hat{a}_2({\bf k})\phi_{\bf k}^{\rm
mcc}(\eta)\Big)e^{i{\bf k}\cdot{\bf x}}+~{\rm h.c.}\Big],
\end{eqnarray}
where two commutation relations take the forms
\begin{eqnarray}
&&\hspace*{-2em}[\hat{a}_i({\bf k}),~\hat{a}_j^{\dag}({\bf
k^{\prime}})]=\left(
  \begin{array}{cc}
   1 & 0  \\
    0 & -1 \\
  \end{array}
 \right)\delta^3({\bf k}-{\bf k}^{\prime}).\label{comrell}
\end{eqnarray}
It is noted that two mode operators ($\hat{a}_1({\bf
k}),\hat{a}_2({\bf k})$) are necessary to take into account of
fourth-order theory quantum mechanically  as the Pais-Uhlenbeck
fourth-order oscillator has been  shown in Ref.~\cite{Mannheim:2004qz}.
In addition, two Wronskian conditions are found to be
\begin{eqnarray}
&&\hspace*{-2em}\Big[\phi_{\bf k}^{\rm mmc}\Big\{\Big(\phi_{\bf
k}^{*\rm mmc}(\eta)\Big)'''++2k^2\Big(\phi_{\bf k}^{*\rm
mmc}(\eta)\Big)'-2aHk^2\phi_{\bf
k}^{*\rm mmc}(\eta)\Big\}\nonumber\\
&&\hspace*{-2em}-\phi_{\bf k}^{\rm mcc}\Big\{\Big(\phi_{\bf k}^{*\rm
mcc}(\eta)\Big)'''+2k^2\Big(\phi_{\bf k}^{*\rm
mmc}(\eta)\Big)'-2aHk^2\phi_{\bf
k}^{*\rm mcc}(\eta)\Big\}\Big]-~c.c.=i,\nonumber\\
&&\hspace*{-2em}\Big[\Big(\phi_{\bf k}^{\rm
mmc}\Big)'\Big\{\Big(\phi_{\bf k}^{*\rm
mmc}(\eta)\Big)''+2aH\Big(\phi_{\bf
k}^{*\rm mmc}(\eta)\Big)'\Big\}\nonumber\\
&&\hspace*{5em}-\Big(\phi_{\bf k}^{\rm
mcc}\Big)'\Big\{\Big(\phi_{\bf k}^{*\rm
mcc}(\eta)\Big)''+2aH\Big(\phi_{\bf k}^{*\rm
mcc}(\eta)\Big)'\Big\}\Big]-c.c.=-i,\label{wconb}
\end{eqnarray}
which will be used  to fix $c_{\rm mmc}$ and $c_{\rm mcc}$ as
\begin{equation}
\phi_{\bf k}^{\rm mmc}=\frac{1}{\sqrt{2^2k^3}}(i+z)e^{iz},~~
\phi_{\bf k}^{\rm
mcc}=\frac{1}{\sqrt{2^2k^3}}ize^{iz}\label{p2sol2}.
\end{equation}
 On the other hand,
the power spectrum~\cite{Baumann:2009ds} of the scalar is defined by
\begin{eqnarray}\label{pow}
\langle0|\hat{\varphi}(\eta,0)\hat{\varphi}(\eta,0)|0\rangle=\int
\frac{dk}{k}{\cal P}_{\varphi}(\eta,k).
\end{eqnarray}
Considering the Bunch-Davies vacuum state imposed by  $\hat{a}_{\bf
k}|0\rangle=0 $ and $\hat{b}_{\bf k}|0\rangle=0$, (\ref{pow}) is
computed  as
\begin{eqnarray}
{\cal P}_{\rm
\varphi}(\eta,k)&=&\frac{k^3}{2\pi^2}\left(\Big|\phi_{\bf
k}^{\rm mmc}\Big|^2-\Big|\phi_{\bf k}^{\rm mcc}\Big|^2\right)\label{powf}\\
&=&\frac{1}{8\pi^2}\Big[(1+z^2)-z^2\Big]\label{powf22}\\
&=&\frac{1}{8\pi^2}\label{powff}
\end{eqnarray}
 Importantly, the minus sign ($-$) in
(\ref{powf}) appears because the unusual commutation relation
($\hat{a}_2({\bf k}),\hat{a}^{\dag}_2({\bf k^{\prime}})$) for ghost
state was used. There is a cancelation between $z^2$ and $-z^2$
thanks to its ghost-like contribution.

Finally, the conformally coupled scalar equation (\ref{eqwi3-2})
from (\ref{WIST3}) leads to the linearized equation around the dS
spacetime as
\begin{equation}
(\Box-2H^2)\varphi^{\rm mcc}=0,
\end{equation}
which is the same equation as in (\ref{meq}). Its normalized
solution takes the form
\begin{equation}
\tilde{\phi}_{\bf k}^{\rm
mcc}=\frac{H}{\sqrt{2k^3}}ize^{iz}\label{p2sol22}.
\end{equation}
In this case, we obviously have a scale-variant scalar power spectrum
\begin{equation} {\cal
P}_{\rm mcc}=\frac{k^3}{2\pi^2}\Big|\tilde{\phi}_{\bf k}^{\rm
mcc}\Big|^2=\Big(\frac{k}{2\pi a}\Big)^2,
\end{equation}
which depends on wave number $k$.

\subsection{Vector power spectrum}

Let us  consider Eq.(\ref{veq}) for vector perturbation and then,
expand  $\Psi_i$ in plane waves with the linearly polarized states
\begin{eqnarray}\label{psim}
\Psi_i(\eta,{\bf x})=\frac{1}{(2\pi)^{\frac{3}{2}}}\int d^3{\bf
k}\sum_{s=1,2}p_i^{s}({\bf k})\Psi_{\bf k}^{s}(\eta)e^{i{\bf
k}\cdot{\bf x}},
\end{eqnarray}
where  $p^{1/2}_{i}$ are linear polarization   vectors with
 $p^{1/2}_i
p^{1/2, i}=1$.  Also, $\Psi_{\bf k}^{s}$ denote linearly polarized
vector modes. Plugging (\ref{psim}) into the equation (\ref{veq}),
one finds the equation
\begin{eqnarray}\label{v0eq}
\Big[\frac{d^2}{d\eta^2}+k^2\Big]\Psi_{\bf k}^s(\eta)=0.
\end{eqnarray}
Introducing  $z=-k\eta$, Eq.(\ref{v0eq}) takes a simple form
\begin{equation}\label{v1eq}
\Big[\frac{d^2}{dz^2}+1\Big]\Psi_{\bf k}^{s}(z)=0\end{equation}
whose  solution is given by \begin{equation} \Psi_{\bf k}^{s}(z)\sim
e^{\pm iz}.\end{equation} Here a positive frequency solution is
given by $e^{iz}$.

Now, let us calculate vector power spectrum. For this purpose, we
define a commutation relation for the vector. In the bilinear action
(\ref{vpeq}), the momentum for the field $\Psi_j$ is found to be
\begin{eqnarray}\label{vconj}
\pi_{\Psi}^{j}=-\frac{\alpha}{2}\nabla^2\Psi'^{j}.
\end{eqnarray}
Note that one observes an unusual factor of Laplacian $\nabla^2$
which reflects that the vector $\Psi_i$ is not a canonically
well-defined vector because it originates from  the fourth-order
conformal gravity. The  quantization is implemented by imposing the
commutation relation
\begin{eqnarray}\label{vcomm}
[\hat{\Psi}_{j}(\eta,{\bf x}),\hat{\pi}_{\Psi}^{j}(\eta,{\bf
x}^{\prime})]=2i\delta({\bf x}-{\bf x}^{\prime})
\end{eqnarray}
with $\hbar=1$. Then, the operator $\hat{\Psi}_{j}$ can be expanded
in Fourier modes as
\begin{eqnarray}\label{vex}
\hat{\Psi}_{j}(\eta,{\bf x})=\frac{1}{(2\pi)^{\frac{3}{2}}}\int
d^3{\bf k}\sum_{s=1,2}\Big(p_{j}^{s}({\bf k})\hat{a}_{\bf
k}^{s}\Psi_{\bf k}^{s}(\eta)e^{i{\bf k}\cdot{\bf x}}+{\rm h.c.}\Big)
\end{eqnarray}
and the operator $\hat{\pi}_{\Psi}^{j}=\frac{\alpha
k^2}{2}\hat{\Psi}'^{j}$ can be obtained from (\ref{vex}). Plugging
(\ref{vex}) and $\hat{\pi}_{\Psi}^{j}$ into (\ref{vcomm}), we find
the commutation relation and Wronskian condition for normalization
as
\begin{eqnarray}
&&\hspace*{-2em}[\hat{a}_{\bf k}^{s},\hat{a}_{\bf k^{\prime}}^{
s^{\prime}\dag}]=\delta^{ss^{\prime}}\delta^3({\bf k}-{\bf
k}^{\prime}),\label{comm0}\\
&&\hspace*{-2em}\Psi_{\bf k}^{s}\Big(\frac{\alpha
k^2}{2}\Big)(\Psi_{\bf k}^{*s})^{\prime}-{\rm c.c.}=i \to \Psi_{\bf
k}^{s}\frac{d\Psi_{\bf k}^{*s}}{dz}-{\rm c.c.}=-\frac{2i}{\alpha
k^3}. \label{vwcon}
\end{eqnarray}
  We
choose the  positive frequency mode  normalized by the Wronskian
condition
\begin{eqnarray} \label{vecsol}
\Psi_{\bf k}^{s}(z) =\sqrt{\frac{1}{\alpha k^3}} e^{iz}
\end{eqnarray}
as the solution to (\ref{v1eq}).
 On the other hand, the
vector power spectrum is defined by
\begin{eqnarray}\label{powerv}
\langle0|\hat{\Psi}_{j}(\eta,0)\hat{\Psi}^{j}(\eta,0)|0\rangle=\int
\frac{dk}{k}{\cal P}_{\Psi}(\eta,k),
\end{eqnarray}
where we take  the Bunch-Davies vacuum $|0\rangle$  by imposing
$\hat{a}_{\bf k}^{s}|0\rangle=0$. The vector  power spectrum ${\cal
P}_{\Psi}$ leads to   \begin{equation} \label{vecpt}{\cal
P}_{\Psi}\equiv\sum_{s=1,2}\frac{k^3}{2\pi^2}\Big|\Psi_{\bf
k}^{s}\Big|^2.\end{equation}
 Plugging (\ref{vecsol}) into (\ref{vecpt}), we find a scale-invariant  power  spectrum for a vector
perturbation
\begin{eqnarray} \label{vec-pow}
{\cal P}_{\Psi}=\frac{1}{\pi^2\alpha^2}.
\end{eqnarray}

\subsection{Tensor power spectrum}

Now, let us take   Eq.(\ref{heq}) to compute  tensor power spectrum. In this
case, the metric tensor $h_{ij}$ can be expanded in Fourier modes
\begin{eqnarray}\label{hijm}
h_{ij}(\eta,{\bf x})=\frac{1}{(2\pi)^{\frac{3}{2}}}\int d^3{\bf
k}\sum_{s={\rm +,\times}}p_{ij}^{s}({\bf k})h_{\bf
k}^{s}(\eta)e^{i{\bf k}\cdot{\bf x}},
\end{eqnarray}
where  $p^{s}_{ij}$ linear polarization tensors with $p^{s}_{ij}
p^{s,ij}=1$. Also, $h_{\bf k}^{s}(\eta)$ represent linearly
polarized tensor modes. Plugging (\ref{hijm}) into (\ref{heq}) leads
to the fourth-order differential equation
\begin{eqnarray}
&&(h_{\bf k}^{s})^{''''}+2k^2(h_{\bf k}^{s})^{''}+k^4h_{\bf k}^{s}
=0,\label{heq2}
\end{eqnarray}
which is further rewritten as a factorized form
\begin{eqnarray}
&&\Bigg[\frac{d^2}{d\eta^2}+k^2\Bigg]^2h_{\bf k}^{s}(\eta)=0.
\label{hee}
\end{eqnarray}
Introducing  $z=-k\eta$, Eq.(\ref{hee}) can be rewritten as  a degenerate
fourth-order equation,
\begin{equation}\label{hc0}
\Big[\frac{d^2}{dz^2}+1\Big]^2h_{\bf k}^{s}(z)=0.\end{equation} This
is  the same equation as for a degenerate Pais-Uhlenbeck (PU)
oscillator~\cite{Mannheim:2004qz} and its solution is given by
\begin{equation} \label{desol}
h_{\bf k}^{s}(z)=\frac{N}{2k^2}\Big[(a_2^s+a_1^s z)e^{iz}+c.c.\Big]
\end{equation}
 with $N$ the normalization constant. After quantization, $a^s_2$
 and $a^s_1$ are promoted to operators $\hat{a}^s_2({\bf k})$ and $\hat{a}^s_1({\bf
 k})$ $(h_{\bf k}^{s}\to\hat{ h}_{\bf k}^{s}$).   The presence of $z$
in $(\cdots)$ reflects clearly that $ h_{\bf k}^{s}(z)$ is a
solution to the degenerate
 equation (\ref{hc0}).
 Together with $N=\sqrt{2/\alpha}$, the
canonical quantization could be
 accomplished by introducing
  commutation
relations between $\hat{a}_i^s({\bf k})$ and
$\hat{a}^{s\dagger}_j({\bf k}')$ as~\cite{Kim:2013mfa}
 \begin{equation} \label{scft}
 [\hat{a}_i^s({\bf k}), \hat{a}^{s^{\prime}\dagger}_j({\bf k}')]= 2k \delta^{ss'}
 \left(
  \begin{array}{cc}
   0 & -i  \\
    i & 1 \\
  \end{array}
 \right)\delta^3({\bf k}-{\bf k}').
 \end{equation}
On the other hand, the tensor power spectrum is  defined by
\begin{eqnarray}\label{power}
\langle0|\hat{h}_{ij}(\eta,0)\hat{h}^{ij}(\eta,0)|0\rangle=\int
\frac{dk}{k}{\cal P}_{\rm h}(\eta,k).
\end{eqnarray}
Here we choose the Bunch-Davies vacuum $|0\rangle$ by imposing
$\hat{a}_i^s({\bf k})|0\rangle=0$ for $i=1,2$. Using the definition
\begin{equation} \label{def-tps}
{\cal P}_{{\rm h}}\equiv \sum_{s,s'={+,\times}}{\cal P}^{ss'}_{\rm
h} \end{equation}
 and
substituting  $\hat{h}_{ij}(\eta,0)$ together with (\ref{scft}) into
(\ref{power}), then one finds that ${\cal P}^{ss'}_{\rm h}$ is given
by
\begin{eqnarray}
{\cal P}^{ss'}_{\rm h}&=&\frac{k}{4\pi^2\alpha^2}\int d^3{\bf
k}'\Big[\frac{1}{k'^2}p_{ij}^{s}({\bf k})p^{ijs'}({\bf k}')
\times\langle0|\Big([\hat{a}_2^s({\bf k}), \hat{a}^{s^{\prime}\dagger}_2({\bf k}')]+z[\hat{a}_2^s({\bf k}), \hat{a}^{s^{\prime}\dagger}_1({\bf k}')]\nonumber\\
&&\hspace*{7em}+z[\hat{a}_1^s({\bf k}),
\hat{a}^{s^{\prime}\dagger}_2({\bf k}')]+z^2[\hat{a}_1^s({\bf k}),
\hat{a}^{s^{\prime}\dagger}_1({\bf k}')]\Big)
|0\rangle\Big]\label{spss}\\
&=&\frac{1}{2\pi^2\alpha^2}p_{ij}^{s}p^{ijs'}\delta^{ss'}[1-iz+iz+0\times
z^2]\label{pss}.
\end{eqnarray}
In obtaining (\ref{pss}), we used the commutation relations of
(\ref{scft}) which reflect the quantum nature of Weyl-invariant
tensor theory like a degenerate PU oscillator. A cancelation between
$iz$ and $-iz$ occurs, showing that this is slightly different from
that between $z^2$ and $-z^2$ in (\ref{powf22}) for the
Weyl-invariant scalar theory of  a non-degenerate PU oscillator.
Finally, from (\ref{def-tps}) and (\ref{pss}),  we obtain a scale-invariant tensor
power spectrum
\begin{equation} \label{tensorp11}
{\cal P}_{\rm h}=\frac{1}{\pi^2\alpha^2},
\end{equation}
which is the same form as for the vector power spectrum
(\ref{vec-pow}).

\section{Discussions}
First of all, we have emphasized  a deep connection between
Weyl-invariance of action (fourth-order theory) and scale-invariance
of power spectrum in dS spacetime.

In deriving the power spectra, we have used  two different
quantization schemes for fourth-order theory: non-degenerate PU
oscillator was employed for quantizing Weyl-invariant scalar theory,
while  degenerate PU oscillator could be used to quantize
Weyl-invariant tensor theory of conformal gravity.

However, we have found the same ambiguity of power spectra in dS
spacetime such that the scalar power spectrum takes the form $\to
1/2(2\pi)^2$ instead of the conventional spectrum
$(H/2\pi)^2[1+(k/aH)^2]$ for a second-order scalar theory of
massless scalar and the tensor power spectrum is given by $1/\pi^2
\alpha^2$ instead of $2(H/\pi M_{\rm P})^2[1+(k/aH)^2]$ for a
second-order tensor theory (Einstein gravity) of massless graviton.
Here $H^2$ was missed and there is no way to restore it in this
approach.   If one has used the Krein space quantization which is
the generalization of the Hilbert space to quantize a massless
scalar in dS space~\cite{Mohsenzadeh:2013bba}, its power spectrum
has led to $(H/2\pi)^2$ which is also scale-invariant as a result of
elimination of scale-dependent term of $(k/2\pi a)^2$.  However,
this method to derive a scale-invariant scalar spectrum is an ad hoc
approach because it has dealt with a second-order scalar theory.

Now, we ask whether our model (\ref{WIST}) is just a toy model for
providing scale-invariant power spectra of scalar and tensor fields
or has really an application to the early stage of the universe
(inflation).  We remind the reader that  power spectra have been
computed based on the dS spacetime (eternal inflation).  However, a
slow-roll inflation is quasi-dS spacetime with graceful exit. In the
slow-roll inflation, the scale-dependence of power spectra appears
when fluctuations of scalar and tensor exit the comoving Hubble
radius [$1/(aH)$] even for choosing the superhorizon limit of
$z=k/aH\to 0$~\cite{Baumann:2009ds}. It seems difficult  to  compute
power spectra of scalar and tensor when one takes a slow-roll
inflation.
 Hence,  our model (\ref{WIST}) is suggested to be a toy model for providing
scale-invariant power spectra of scalar and tensor fields in dS
inflation, which are independent of scale $z(k)$ in whole range of
$z$.

At this stage, we would like to comment on two different
Weyl-invariant theories. The action $S_{\rm ST1}$ in Eq.(\ref{WIST})
gives us a  scale-invariant scalar spectrum of $1/2(2\pi)^2$, while
the action $S_{\rm ST3}$ in Eq.(\ref{WIST3}) provides a
scale-variant scalar spectrum of $(k/2\pi a)^2$. The former scalar
has dimensionless  (Weyl-invariant), whereas the latter has
dimension $1$ (Weyl-variant).

Finally, we should mention the ghost issues (negative-norm state)
because Weyl-invariant scalar-tensor theory is a fourth-order
scalar-tensor theory. In general, a fourth-order scalar theory
implies two second-order theories with opposite signs in
diagonalized commutation relations (\ref{comrell}), while a
degenerate fourth-order tensor theory implies two second-order
tensor theories with non-diagonalized commutation relations
(\ref{scft}). In the Weyl-invariant scalar theory, there is
cancelation between $z^2$ (positive-norm state from mmc) and $-z^2$
(negative-norm state from mcc scalar) in the power spectrum
(\ref{powf22}). This reflects the quantization of the non-degenerate
PU oscillator. On the other hand, in the Weyl-invariant tensor
theory, a cancelation between $iz$ and $-iz$ in the power spectrum
(\ref{pss}) occurs as in the quantization scheme of degenerate PU
oscillator. Consequently, there are  no negative-norm states in the
scalar and tensor power spectra in dS spacetime. This may indicate
that the Weyl-invariance forbids the ghost states of  the power
spectra in dS spacetime. As  counter examples, one did not obtain
positive scalar (tensor) power spectrum from  the nonminimal
derivative coupling with fourth-order term~\cite{Myung:2015xha}
(Einstein-Weyl gravity~\cite{Deruelle:2012xv,Myung:2014jha}), which
are not obviously Weyl-invariant.

 \vspace{0.25cm}

 {\bf Acknowledgement}

\vspace{0.25cm}
This work was supported by the 2015 Inje University
Research Grant.

\end{document}